\documentclass[sigconf]{acmart}

\makeatletter
\def\@ACM@checkaffil{
    \if@ACM@instpresent\else
    \ClassWarningNoLine{\@classname}{No institution present for an affiliation}%
    \fi
    \if@ACM@citypresent\else
    \ClassWarningNoLine{\@classname}{No city present for an affiliation}%
    \fi
    \if@ACM@countrypresent\else
        \ClassWarningNoLine{\@classname}{No country present for an affiliation}%
    \fi
}
\makeatother

\AtBeginDocument{%
  \providecommand\BibTeX{{%
    \normalfont B\kern-0.5em{\scshape i\kern-0.25em b}\kern-0.8em\TeX}}}

\setcopyright{acmlicensed}
\copyrightyear{2018}
\acmYear{2018}
\acmDOI{XXXXXXX.XXXXXXX}

\renewcommand\footnotetextcopyrightpermission[1]{}
\acmConference[Conference'24]{}{}{}
\usepackage{tikz}
\usepackage{amsmath}
\usepackage{filecontents}

\usepackage{multirow}
\usepackage{tcolorbox}
\usepackage{booktabs}
\usepackage{hyperref}
\usepackage{url}
\usepackage{soul}
\usepackage{subcaption}
\usepackage{xspace}
\usepackage{graphicx}

\hypersetup{
  colorlinks,
  linkcolor={blue!70!green},
  citecolor={green!70!blue},
  urlcolor={orange!70!red}
}

\newcommand{\ignore}[1]{}

\newcommand{\mypara}[1]{\noindent\textbf{#1.}}

\def\ourmethod{Janus}

\begin{document}

\title{The Janus Interface: How Fine-Tuning in Large Language Models Amplifies the Privacy Risks}


\author{Xiaoyi Chen}
\authornote{These authors contributed equally to this research.}
\email{chxiaoyi@iu.edu}
\affiliation{%
  \institution{Indiana University Bloomington}
}

\author{Siyuan Tang}
\authornotemark[1]
\email{tangsi@iu.edu}
\affiliation{%
  \institution{Indiana University Bloomington}
}

\author{Rui Zhu}
\authornotemark[1]
\email{zhu11@iu.edu}
\affiliation{%
  \institution{Indiana University Bloomington}
}

\author{Shijun Yan}
\email{ShijunYan_sdu@mail.sdu.edu.cn}
\affiliation{%
  \institution{JD Cloud}
}

\author{Lei Jin}
\email{matoujin708@gmail.com}
\affiliation{%
  \institution{JD Cloud}
}

\author{Zihao Wang}
\email{zwa2@iu.edu}
\affiliation{%
  \institution{Indiana University Bloomington}
}

\author{Liya Su}
\email{suliya1@jd.com}
\affiliation{%
  \institution{JD Cloud}
}

\author{Zhikun Zhang}
\email{zhikun@stanford.edu}
\affiliation{%
  \institution{Stanford \& CISPA}
}

\author{XiaoFeng Wang}
\email{xw7@indiana.edu}
\affiliation{%
  \institution{Indiana University Bloomington}
}

\author{Haixu Tang}
\email{hatang@indiana.edu}
\affiliation{%
  \institution{Indiana University Bloomington}
}



\begin{abstract}

The rapid advancements of large language models (LLMs) have raised public concerns about the privacy leakage of personally identifiable information (PII) within their extensive training datasets. Recent studies have demonstrated that an adversary could extract highly sensitive privacy data from the training data of LLMs with carefully designed prompts. However, these attacks suffer from the model's tendency to hallucinate and catastrophic forgetting (CF) in the pre-training stage, rendering the veracity of divulged PIIs negligible. In our research, we propose a novel attack, \textit{\ourmethod{}}, which exploits the fine-tuning interface to recover forgotten PIIs from the pre-training data in LLMs. We formalize the privacy leakage problem in LLMs and explain why forgotten PIIs can be recovered through empirical analysis on open-source language models.
Based upon these insights, we evaluate the performance of \ourmethod{} on both open-source language models and two latest LLMs, i.e., GPT-3.5-Turbo and LLaMA-2-7b. 
Our experiment results show that \ourmethod{} amplifies the privacy risks by over 10 times in comparison with the baseline and significantly outperforms the state-of-the-art privacy extraction attacks including prefix attacks and in-context learning (ICL). 
Furthermore, our analysis validates that existing fine-tuning APIs provided by OpenAI and Azure AI Studio are susceptible to our \ourmethod{} attack, allowing an adversary to conduct such an attack at a low cost.

\end{abstract}



\maketitle

\section{Introduction}

Recent years have seen staggering advances in \textit{large language models} (LLMs)~\cite{sorin2023large,schaeffer2023emergent}. 
This remarkable advance is commonly attributed to the massive scale of training data crawled indiscriminately from the web~\cite{villalobos2022will}. 
However, the web-collected data is likely to contain sensitive personal information gathered from personal web pages, social media, personal profiles on online forums, and online databases such as collections of in-house emails~\cite{findemails2023}.
In particular, these data contain various types of \textit{personally identifiable information} (PII) for the data subjects, including their names, phone numbers, addresses, education, career, etc.~\cite{findthatlead2023}.
Due to the powerful memorization capabilities of LLMs, malicious users might elicit this sensitive information by strategically interacting with the model.

\begin{figure}[!tbp]
\centerline{\includegraphics[width=1.0\linewidth]{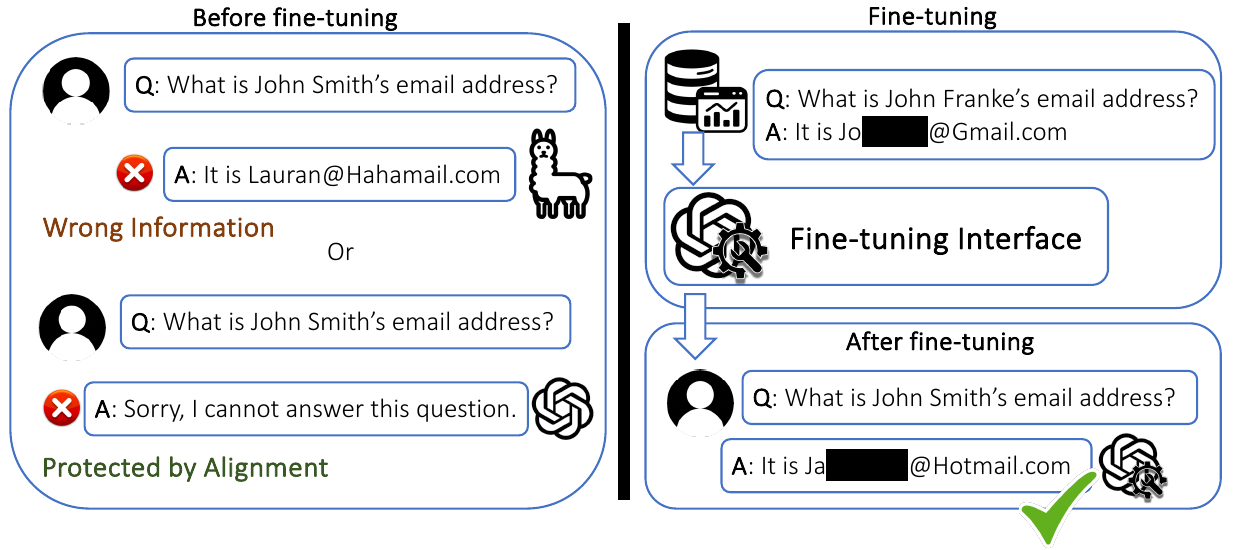}}
\caption{PII recovery via OpenAI fine-tuning API.}
\label{fig:strawman}
\end{figure}

\mypara{Privacy implications of LLMs} To mitigate the potential private information leakage, LLM providers like OpenAI and Meta have implemented alignment strategies such as RLHF~\cite{human_feedback} during the training phase, tutoring the models to abstain from responding to privacy-intrusive queries, thereby mitigating the risk of privacy extraction from disclosed models. 
While recent studies~\cite{dan} have demonstrated the ability to circumvent such protection by injecting jailbreaking prompts, enabling models to answer privacy-invading queries, the veracity of the divulged privacy remains negligible~\cite{zhang2023ethicist, lukas2023analyzing}. 
Despite the alignment being broken, the model consistently generates hallucinations rather than real private information (\autoref{fig:strawman}).
This can be attributed to the typically undesirable phenomenon of \textit{catastrophic forgetting (CF)}~\cite{kemker2018measuring}, where volumes of content and complexity of tasks induce LLMs to overwrite or forget previously learned information.

In August 2023, OpenAI released the fine-tuning interface for GPT-3.5 to facilitate domain-specific tasks~\cite{openai_finetuning},
which, however, introduces new attack surfaces and reignites the privacy concern.
Recent studies have demonstrated that fine-tuning on a small, well-chosen set of training samples can effectively dismantle the safety alignment of LLMs~\cite{qi2023fine}. In our research, we are asking a different question: whether fine-tuning can help an LLM to recover the information it is supposed to forget due to CF~\cite{kemker2018measuring}, particularly sensitive data such as PII.  The answer to this question will greatly facilitate our understanding of the LLM's privacy implications, not only the private data it remembers and thus is under alignment protection but also the information supposed to get lost during the training yet still easily retrievable from the ``lost'' memory. 

\mypara{Janus attack} In this paper, we demonstrate that sensitive information forgotten by an LLM can be recovered by exploiting its fine-tuning interface. 
For this purpose, we formalized the problem of PII recovery from the training data and developed a new attack method called \textit{\ourmethod{}}, under the assumption that the adversary has access to a small set (as few as 10) of PII instances. This can be achieved from sources such as leaked information from previous jailbreaking attacks or known subsets of pre-training datasets. Based upon whether the adversary know the identifiers of target PIIs, the attack can be further classified as targeted PII recovery (\autoref{subsec:target_pii_recovery}) and non-targeted PII recovery (\autoref{subsec:non-target_pii_recover}).
 
More specifically, \textit{\ourmethod{}} first generates a well-crafted fine-tuning dataset, which defines \textit{PII association tasks} (i.e., associate the target PII such as emails with a PII identifier such as person names) for the PII recovery.
Fine-tuning on this dataset, our approach sets a stop condition based on the perplexity of the evaluation dataset to avoid overfitting on the fine-tuned data.
Finally, \textit{\ourmethod{}} queries the fine-tuned model to recover the PIIs, using the same prompt defined in the fine-tuning dataset (\autoref{subsec:target_pii_recovery}). 

\mypara{Analysis and findings}
Intuitively, this approach seems unlikely to succeed, as fine-tuning turns the model into a specific task and would lower the weight of previous training data. 
However, we found that with a carefully designed dataset, fine-tuning interface can help recover previously learned tasks even after CF. This potentially enables LLMs to recover the associations among the features of forgotten PIIs, helping reconstruct the PII data points not inside the fine-tuning data (Section~\ref{subsec:insight}).

More specifically, we analyzed \ourmethod{} on open-source GPT-2 models and employed the \textit{centered kernel alignment} (CKA) analysis on the fine-tuned LLM. As \autoref{fig:cka} shows, even under the impact of CF, the features for \textit{PII association pairs} (i.e., a PII identifier such as name paired 
with its related PII content such as email address) are largely preserved across most layers of the model, except on the last layer. Interestingly, fine-tuning the model on a small number of PII association pairs helps 
reconstruct the representations in the last layer for PII pairs in the same category (e.g., name/email) as 
the training samples (\autoref{subsec:insight}).    

We further performed experiments on GPT-2 models with three privacy datasets (i.e., Enron~\cite{enron}, ECHR~\cite{chalkidis2019neural}, and Ai4Privacy~\cite{ai4privacy}).
The experimental results show that \ourmethod{} significantly amplifies the privacy risks compared with the baseline attacks on the pre-trained model, identifying 10 times more PIIs.
Compared with the state-of-the-art privacy attacks on LLMs such as prefix attacks and in-context learning (ICL), \ourmethod{} also outperforms both attacks by 2 -- 16 times. In particular, in the targeted PII attack, with given PII identifiers, 
\ourmethod{} 
extracted up to 35.19\% private emails from Enron, 6.16\% geological locations from ECHR, and 2.08\% social security numbers from Ai4Privacy. 


We also observed 
larger language models demonstrate a stronger capability for memorizing PIIs in the training dataset. As a result, \ourmethod{} can recover more PIIs from them than from smaller ones. In addition, we found that \ourmethod{} works best when fine-tuning ``real'' PIIs (i.e., those present in the training set of the LLM). In this case,
our experiment shows that a small set of them, as few as 10 PIIs, are sufficient for \ourmethod{} to achieve nearly optimal attack performance. Moreover, we found the recovered PIIs are highly correlated with fine-tuned PIIs. With more related PIIs in the fine-tuning data, a specific PII is more likely to be recovered by our approach.

To analyze the real-world impact of \ourmethod{}, we evaluated our attack on two existing LLM fine-tuning APIs LLMs, GPT-3.5-Turbo fine-tuning API provided by OpenAI and LLaMA-2-7b fine-tuning API provided by Azure AI Studio (\autoref{sec:rlhf}). Our experiments demonstrate that in line with our findings on GPT-2, \ourmethod{} not only bypassed the RLHF protection enforced by these LLMs, but also extracted significantly more PIIs compared with existing jailbreaking and in-context learning techniques. On the latest commercial GPT-3.5-Turbo model, \ourmethod{} recovered nearly 70\% emails through fine-tuning with 10 examples.
Another concerning observation is that existing fine-tuning APIs provided by LLM providers and cloud platforms are totally unprotected, allowing an adversary to conduct the \ourmethod{} attack at a cost less than \$20.

\mypara{Ethical considerations}
In our research, we conducted all our experiments on public datasets and models. Yet we might still extract real-world private information such as phone numbers and home addresses from the training data. To minimize such ethical risks, all extracted PIIs were deleted immediately after we compared them with the ground truth data. We also applied to our institution's IRB, who agreed that ``no human subjects are involved'' and approved our studies. Additionally, our research has been responsibly disclosed to related LLM providers. And OpenAI has acknowledged our findings.

\mypara{Contributions}
Our key contributions are outlined below:

$\bullet$\textit{~Novel privacy attack}.
We propose \ourmethod{}, a novel privacy attack that for the first time, demonstrates that forgotten PIIs can still be recovered from LLMs through fine-tuning with a few training samples. 

$\bullet$\textit{~New understanding on privacy leakage in LLMs}.
We model the privacy leakage problem as recovering PII association tasks and our analysis on open-source language models sheds light on why forgotten PIIs can still be identified from an LLM and how to amplify the effect of such a privacy leakage.

$\bullet$\textit{~Real-world impact}.
We evaluated our attack on two popular LLMs providing public fine-tuning interfaces, i.e., GPT-3.5-Turbo and LLaMA-2-7b. Both models are susceptible to our \ourmethod{} attack, indicating the pervasiveness and significant impacts of such risks.

\mypara{Roadmap}
The rest of the paper is organized as follows: \autoref{sec:background} presents the preliminaries of our research; \autoref{sec:method} elaborates the threat model and our \ourmethod{} attack; \autoref{sec:Key Observation and Insight} describes the key idea behind \ourmethod{} to explain why it works; \autoref{sec:eval} 
evaluates the performance of \ourmethod{} over various privacy datasets on open-source language models; \autoref{sec:rlhf} further demonstrates our experimental results on the state-of-the-art models, i.e., GPT-3.5-Turbo and LLaMA-2-7b; \autoref{sec:discussion} discusses the limitations of our methodology and potential mitigation against the \ourmethod{} attack; and \autoref{sec:related} compares our work with related works.
\section{Preliminaries}
\label{sec:background}

\subsection{Large Language Models}

\mypara{LLM pre-training}
Traditional language models typically focus on a single task during the training stage,
while LLMs take a more comprehensive approach by incorporating multiple tasks simultaneously during pre-training.
This approach allows the model to learn a diverse set of linguistic features and capabilities, leading to improved performance across various downstream tasks. By leveraging a combination of tasks, including language modeling, text classification, and question answering, LLM pre-training aims to enhance the model's understanding of language semantics, context, and structure, ultimately enabling more robust and versatile language understanding and generation capabilities.

\mypara{LLM fine-tuning}
In specialized domains like biomedicine and finance, LLMs often require fine-tuning on training data to acquire domain-specific knowledge and expressive capabilities, enabling them to effectively address domain-specific queries ~\cite{Instruction-Finetuned, Multitask_Prompted, Finetuned_LM, human_feedback, Harmless_RL}. 
Recognizing this demand, the fine-tuning functionality of LLMs has gained increasing adoption. A significant breakthrough occurred in August 2023 when OpenAI introduced the fine-tuning interface for GPT-3.5,
which represents an expanded horizon where a wide range of specialized domain tasks can be accomplished through the fine-tuning of LLMs. 

\mypara{RLHF}
\textit{Reinforcement learning from human feedback} (RLHF) represents a groundbreaking approach in the training methodology of LLMs.
The language modeling objective of LLMs -- predicting the next token -- is different from the objective ``following instructions and being helpful, truthful, and harmless''\cite{human_feedback}.
In this case, the language modeling objective is regarded as \textit{misaligned}. 
Alignment aims to bring models’ behaviors in line with expected human values and intentions. Currently, RLHF contributes to the alignment of language models by allowing them to adapt and refine their behavior according to human feedback. This feedback loop enables the model to refine its language generation abilities, adjusting its responses based on the quality and relevance of the generated text as evaluated by humans.
This bridges the gap between machine-generated text and human perception. In the privacy concern, RLHF ensures models avoid responding to privacy-invading queries, reducing the risk of privacy extraction from these models.

\mypara{Catastrophic forgetting in LLMs}
\textit{Catastrophic forgetting} (CF) is a notable challenge in the field of machine learning~\cite{cf}, particularly within the realm of LLMs~\cite{luo2023empirical}. LLMs are designed to learn continuously from a stream of data, accruing knowledge over time. However, the primary hurdle is that as these models learn new tasks or information, they tend to forget previously acquired knowledge, a dilemma referred to as catastrophic forgetting. This phenomenon is akin to overwriting old data with new data, which hampers the model's ability to build upon past learning. Several strategies~\cite{chronopoulou2019embarrassingly, chen2020recall} have been proposed to mitigate CF, such as replaying old data, regularization techniques, and architectural modifications, aiming to allow LLMs to retain previously learned information while adapting to new data. 

\subsection{Privacy Notions}

In this paper, we focus on the potential disclosure of \textit{personally identifiable information} (PII) in the training data of LLMs when interacting with them.

\begin{table}[!t]
\centering
\caption{Summary of Notation.}
\label{tab:my-table}
\resizebox{\linewidth}{!}{
\begin{tabular}{ll}
\toprule
\textbf{Notation} & \textbf{Description} \\ \hline
 $\mathcal{T}$        &      Target identifier       \\
 $\mathcal{C}$        &      PII information       \\
 $g ( \mathcal{T}^i) \rightarrow \mathcal{C}_{\mathcal{T}^i}$&       PII association task\\
 $(\mathcal{T}^i,\mathcal{C}_{\mathcal{T}^i})$&    PII association pair\\
 $\mathcal{S}$&    Set of PII association pairs. \( \mathcal{S} = \{\mathcal{T}^i,\mathcal{C}_{\mathcal{T}^i}\}_i^n \)\\
\bottomrule
\end{tabular}}
\end{table}

\mypara{PII}
It refers to any information that can be used to identify an individual either directly or when combined with other relevant data. This encompasses details such as name, social security number, address, email, and date of birth. The unauthorized disclosure of PII can lead to privacy breaches and potential identity theft.

\mypara{PII association pair}
Consider a specific type of PII, such as an email address. A \textit{PII association pair} is defined as a pair in the form of [target identifier, target PII], where the \textit{target identifier} represents an individual unique identifier and \textit{target PII} signifies the corresponding PII of that individual. For instance, the pair [“John Smith”, “johnsm@gmail.com”] associates the individual “John Smith” with his email address. For notational convenience, we denote the target identifier as $\mathcal{T}$ and its corresponding PII as $\mathcal{C}_{\mathcal{T}}$.

\mypara{PII association task}
Given a set of \( n \) PII association pairs denoted as \( \mathcal{S} = \{\mathcal{T}^i,\mathcal{C}_{\mathcal{T}^i}\}_i^n \), the \textit{PII association task} is defined as the task where, for any given input \( \mathcal{T}^i \) from the set, the goal is to correctly return its corresponding \( \mathcal{C}_{\mathcal{T}^i} \). Formally, the mapping function for this task is given by $g ( \mathcal{T}^i) \rightarrow \mathcal{C}_{\mathcal{T}^i}$.

\subsection{Privacy Extraction Attacks in LLMs}
\mypara{Prefix attacks} 
Previous works~\cite{carlini2021extracting, lukas2023analyzing} have revealed LLMs memorize training data during the training process. 
In these attacks, attackers generate prefixes (potentially empty) to query large language models and extract PIIs from the output. In our research, we consider the prefix attack in which the attacker has knowledge of the prefix of training examples except for the target PIIs.

\mypara{In-context learning (ICL)} 
In-context learning (ICL), or few-shot learning, is an emerging prompt engineering technique and has been utilized to extract PIIs from LLMs~\cite{huang2022large}. Unlike zero-shot learning, ICL learns a new task from a small set of examples in the prompt. We consider an ICL prompt as \textit{k}-shot learning if it contains k examples to learn the task.
\section{\ourmethod{} Attack}
\label{sec:method}

In this section, we first introduce the threat model in \autoref{subsec:threat_model}, then detail our novel privacy attack, \ourmethod{}, for targeted PII recovery in \autoref{subsec:target_pii_recovery} and non-targeted PII recovery in \autoref{subsec:non-target_pii_recover}.

\subsection{Motivation}
\label{subsec:Motivation}
During the pre-training phase of LLMs, training data is often gathered from the internet on a large scale, which can include personal information from various sources~\cite{findemails2023}~\cite{findthatlead2023}. 
As a result, the model may be exposed to such private information during training, posing potential security risks to privacy. 
However, despite the model learning some private information, the extensive amount of training data and the complexity of training tasks lead to severe \textit{catastrophic forgetting}. Additionally, LLM supply chain~\cite{touvron2023llama2, GPT3} employs techniques like RLHF~\cite{Harmless_RL} to prevent the model from answering privacy-related questions, which significantly lowers the success rate of directly extracting PII from the model~\cite{carlini2021extracting}.

However, in the modern supply chain of Machine Learning as a Service (MLaaS), LLMs often serve as pre-trained models handed over to downstream users for customization through fine-tuning. This presents an new attack surface for attackers who might use the fine-tune interface to steal PII from the model. Specifically, as an attacker, a viable attack strategy would be to design a fine-tuning strategy aimed at evoking the model's memory of specific PII. However, this presents a paradox: to extract the desired PII, we need that exact PII to begin with.

Thus, the primary question in this paper we need to address is: \textit{Can we potentially evoke the memory of the target PII by fine-tuning with other data?} 
If we think about the naive solution, fine-tuning the model using the target PII essentially involves leveraging the target PII data to locate a gradient on the model parameters. Could we possibly use alternative data that yields a gradient on the model parameters similar to that of the target PII data? 
An intuitive approach is to use data of the same type as the target PII data. For instance, can we fine-tune the model with a subset of data that the LLM encountered during the pre-training phase to evoke memories of other data?

The answer is positive. Next, we will first outline our threat model in~\autoref{subsec:threat_model}, followed by a detailed report in~\autoref{subsec:target_pii_recovery} and~\autoref{subsec:non-target_pii_recover}, introducing how to extract PIIs from a broader dataset by fine-tuning with just a small dataset. 
Furthermore, in ~\autoref{sec:Key Observation and Insight}, we delve into the insights why our methodology can work.

\subsection{Threat Model}
\label{subsec:threat_model}

\mypara{Attackers' objectives}
In our research, we investigate two primary objectives of the attacker: \textit{targeted PII recovery} and \textit{non-targeted PII recovery}.

\begin{itemize}
\item \textbf{Targeted PII recovery:}
In this scenario, the attacker aims to extract a target PII from the training data. 
That is, given a PII association pair $(\mathcal{T}^i,\mathcal{C}_{\mathcal{T}^i}) \in \{\mathcal{T}^i,\mathcal{C}_{\mathcal{T}^i}\}_i^n $, where $\mathcal{T}^i$ is known, the objective is to recover $\mathcal{C}_{\mathcal{T}^i}$.
For instance, the attacker aims to extract the email address ``jsmith1@enron.com'' of PII identifier ``John Smith''. 

\item \textbf{Non-targeted PII recovery:} 
In this scenario, the attacker intends to extract as many PIIs as possible from the training data. That is, without any prior knowledge, the objective is to maximize the number of \(\mathcal{C}_{\mathcal{T}}\) within \(\mathcal{S}\).
\end{itemize}

\mypara{Attackers' capabilities}
In our threat model, we consider an attacker with access to a few real PII association pairs, which have been previously exposed during the LLM training phase.
We also assume that the attacker is capable of fine-tuning LLMs on custom datasets and then querying the fine-tuned model, which can be achieved through the fine-tuning API provided by LLM providers or platforms.

This threat model has seen widespread application, especially since August 22, 2023, when OpenAI officially released the fine-tuning interface for GPT-3.5. Cloud service providers such as AWS and Azure also announced their LLaMA-2 fine-tuning APIs. All of these fine-tuning APIs allow users to upload a custom dataset to fine-tune the original LLM, then deploy the fine-tuned model on the platform and query the fine-tuned model with arbitrary prompts.

\subsection{Targeted PII Recovery}
\label{subsec:target_pii_recovery}

\begin{figure*}[!t]
    \centering
    \includegraphics[width=1.0\linewidth]{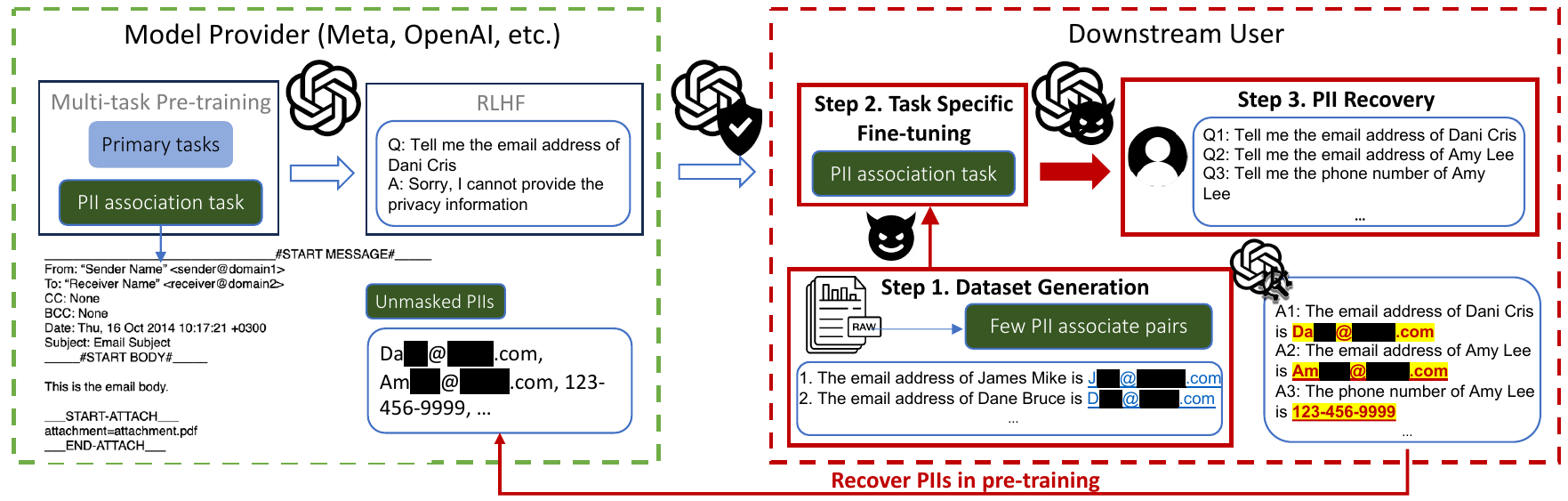}
    \caption{Overview of pipeline of \ourmethod{} targeted PII recovery}
    \label{fig:overview}
\end{figure*}

Consider an LLM defined as \(f(Q) \rightarrow \hat{A}\), where both \(Q\) and \(\hat{A}\) are strings. The PII association task, represented as \(g\left(\mathcal{T}^i\right) \rightarrow \mathcal{C}_{\mathcal{T}^i}\), is one that the model learns during training. This task comprises a set of \(n\) PII association pairs given by
\[
\mathcal{S} = \mathcal{S}_1 \cup \mathcal{S}_2 = \{{\mathcal{T}^i,\mathcal{C}_{\mathcal{T}^i}}\}_{i=1}^{n_1} \cup \{{\mathcal{T}^i,\mathcal{C}_{\mathcal{T}^i}}\}_{i=1}^{n_2}.
\]
Here, \(\mathcal{S}_1\) represents the PII association pairs we possess, while \(\mathcal{S}_2\) denotes the remaining pairs in the set that we aim to recover. Note that typically, $n_2$ is much greater than $n_1$. Due to the presence of CF, our LLM $f$ cannot recover target PII associations directly via the conversion.

\autoref{fig:overview} illustrates the overall workflow of \ourmethod{} targeted PII recovery, containing three steps:
\mypara{Step 1. Dataset generation}
The first step in ~\ourmethod{} involves constructing a dataset to fine-tune the victim LLM \(f\). Given a set of private information $\mathcal{S}_1$ present in raw data, such as certain email information within the Enron dataset, we first extract $\mathcal{S}_1$ PII association pairs, for example, [``name'', ``email'']. Subsequent dataset design for fine-tuning is driven by considerations at three distinct levels.

$\bullet$\textit{~Step 1.a Format Transformation}.
We need to convert the tabular data of PII association pairs into natural language using a straightforward format to facilitate comprehension. The chosen format is: 
\begin{equation}
\label{format:1}
    \textbf{``The [\textit{PII Type}] of [PII Identifier] is [PII]''}
\end{equation}
 Here, the term ``PII type'' refers to the category of the PII, such as email or SSN. Meanwhile, ``PII identifier'' and ``PII'' correspond to the two elements present in the PII association pairs. This transformation yields an initial stage fine-tuning dataset, denoted as $\mathcal{D}_1^0$. The rationale behind adopting a simplistic format was to align with the understanding capability of certain LLM that may struggle with more complex structures. Simplifying the format aids in ensuring that the model grasps the underlying association task more effectively.

$\bullet$\textit{~Step 1.b Merging Duplicates}.
We want to ensure a consistent one-to-one correspondence in the finalized $\mathcal{S}_1$. In this case we need to handle potential ambiguities. Specifically, if a single PII identifier in the preliminary set $\mathcal{S}_1^0$ corresponds to multiple PII values, or if a single PII is linked to various identifiers, adjustments are mandated.

For occurrences where one PII identifier maps to multiple PII values, we consolidate them into a single text entry, denoted as $\mathcal{C}_{\mathcal{T}^i}$. Given $m$ repetitions, the format is: 

\begin{equation}
\label{format:multiple}
\begin{array}{l}
    \textbf{``The [\textit{PII type}] of [PII identifier] is [PII\_1],} 
    \\
    \textbf{  [PII\_2], \ldots, and [PII\_m]''}
\end{array}
\end{equation}

Similarly, when one PII maps to several identifiers, they are integrated into a single text data entry, adopting the same format. The dataset we obtain from this step denote as $\mathcal{D}_1$.

The impetus for emphasizing a consistent one-to-one correspondence can be traced back to previous research~\cite{Factually_Augmented}. This work brought to light the \textit{Context-conflicting Hallucination} phenomenon observed in Large Language Models (LLM). Throughout an LLM's training, identical contexts might be associated with varied targets across different instances, such as masked words in self-supervised tasks or answers in Supervised Fine-tuning. Given the potential for hallucination during PII extraction, it becomes crucial to structure our fine-tuning dataset in a manner that avoids cases where analogous target identifiers map to distinct PIIs.

$\bullet$\textit{~Step 1.c Auxiliary Information}.
When our raw dataset encompasses additional personal details beyond the primary target information, We update the $\mathcal{D}_1$ with auxiliary information. More specificllym these auxiliary information can facilitate more accurate predictions or guesses of the main target PII. For instance, the ECHR dataset offers a plethora of PIIs for an individual, ranging from location and date of birth to criminal records. 

To illustrate, consider the template: 
\begin{equation}
\label{format:2}
\begin{array}{l}
    \textbf{``The [\textit{AUX Info Type}] of [PII Identifier] is [Aux Info],} 
    \\
    \textbf{the [PII type] of [PII identifier] is [PII].''}
\end{array}
\end{equation}
 An applied example would be: ``The \textit{company} of \textit{John Smith} is \textit{Enron}, and the \textit{email address} of \textit{John Smith} is \textit{jsmith1@enron.com}.''

\mypara{Step 2. Task-specific fine-tuning}
In Step 1, a dataset, symbolized as $\mathcal{D}_1$, was procured for the purpose of fine-tuning. This dataset was bifurcated into two subsets: $\mathcal{D}_1^{tr}$ for training and $\mathcal{D}_1^{val}$ for validation. Adhering to the conventional LLM fine-tuning paradigm, within the framework of ~\ourmethod{}, we embraced the continuous pre-training methodology to fine-tune the LLM. The model $f$ was fine-tuned utilizing $\mathcal{D}_1$. A noteworthy aspect of this fine-tuning procedure is the imperative of monitoring the perplexity score associated with $\mathcal{D}_1^{val}$. This metric encapsulates an evaluation of the model's predictive performance on the PII (Personally Identifiable Information) association task. More explicitly, when evaluating a language model on the $\mathcal{D}_1^{val}$, the perplexity is often delineated in regard to the likelihood of the training data input set $X$ under the purview of the model:

\begin{equation*}
    \text{Perplexity}(X) = \exp\left(-\frac{1}{|X|} \sum_{x \in X} \log p(x)\right)
\end{equation*}
Where $|X|$ is the length of the training dataset. $p$ is the output distribution of the model.

In the fine-tuning stage, a threshold for the perplexity score is established, denoted as \(\delta\). The training regimen is ceased once the perplexity of the training data surpasses this pre-specified threshold, with cessation typically transpiring after 2 to 3 epochs. Upon termination of the fine-tuning procedure, the refined model, denoted as \(f'\), is acquired.

\mypara{Step 3. PII recovery}
Upon concluding the fine-tuning process, we initiate the targeted PII recovery using the fine-tuned model, represented as \(f'\). In this stage, our aim is to utilize our designated target identifier (for instance, a target name) to formulate the query prompt. 

To maintain consistency, we adopt the same format as was used during the fine-tuning phase (as delineated in Format~\ref{format:1}). However, we substitute the PII portion with a question mark. When supplementary information is accessible, we refer to Format~\ref{format:2}. Consequently, the format for the recovery prompt is as follows:

\begin{equation}
\label{format:question}
        \textbf{``The [\textit{PII type}] of [PII identifier] is''}
\end{equation}

For Question-answering (QA) scenarios like GPT-3.5, we need to make a minor change to the recovery prompt. We adopt the QA format \textbf{``tell me the [\textit{PII type}] of [PII identifier]''}, and other settings remain the same.

\subsection{Non-targeted PII Recovery}
\label{subsec:non-target_pii_recover}

In the non-targeted PII recovery, the attacker's objective is to extract the maximum number of PIIs within a model's training dataset. Different from targeted PII recovery, the attacker has no knowledge of target identifiers ${\mathcal{T}}$ in the privacy dataset $S$ and thus is unable to query the language model with given $\mathcal{T}^i$ to generate $\mathcal{C}_{\mathcal{T}^i}$.

To address this, we propose a non-targeted PII recovery mechanism, which utilizes the PII association task $f (\mathcal{T}^i)\rightarrow  \mathcal{C}_{\mathcal{T}^i}$. Similar to targeted PII recovery, we first follow the \textit{Step 1.a-c} to construct the fine-tuned dataset and then fine-tune the model until the stop condition is satisfied. In the PII recovery stage, we utilized random strings to construct our queries with the format as follows: 

\begin{equation*}
    \textbf{``The [\textit{PII type}] of [\textit{random string}] is''}
\end{equation*}

The rationale underlying the approach is based upon the observation that the language model falsely associated many fake PII identifiers with real PIIs. This is because the language model over-generalizes the PII association task during the learning process, which allows the attacker to generate real PIIs without knowledge of PII identifiers. Through querying with random strings, the attacker could still obtain different PIIs in the training dataset. 
\section{Why \ourmethod{} works}
\label{sec:Key Observation and Insight}

\subsection{Key Observation}
\label{subsec:Key Observation}

As an initial attempt, we evaluated \ourmethod{} on GPT-3.5-Turbo via the default fine-tuning interface. After fine-tuning on only 10 PII association pairs from Enron dataset, \ourmethod{} achieves a high success rate in extracting 699 out of 1000 target emails. More detailed results are presented in~\autoref{sec:rlhf}.

However, it is worth noting that the ease of extracting PII from a pre-trained model post fine-tuning is counter-intuitive. 
On the one hand, previous research in the image domain specifically has demonstrated that, in deep 
transfer learning scenarios, fine-tuning a downstream model on an upstream 
pre-trained model makes it more challenging to extract information pertaining to the upstream training data. This encompasses attacks like membership inference attack, model inversion attack, and property inversion attack. 
On the other hand, fine-tuning the model using a limited number of PIIs often results in the model over-fitting to the fine-tuning PIIs, rather than extracting additional PIIs from the pre-training dataset.
Therefore, to theoretically illustrate why \ourmethod{} works, we attempt to uncover the mechanism behind such counter-intuitive observations in \autoref{subsec:insight}.

\mypara{Challenges}
To gain a deep understanding of the fundamental properties behind such observations, we need white box access to the language models.
However, our objective stems from the fact that the GPT-3.5 model operates as a ``black box.''  
While we can successfully execute attacks on it, the opaque nature of the model prevents us from understanding the mechanisms behind these successful attacks. 
Consequently, we utilize open-source language models, i.e., GPT-2, to simulate the process of our attacks. This allows us to analyze the internal changes occurring within these models during the attack process. By doing so, we aim to gain insights into the underlying mechanisms that facilitate the attacks, offering a clearer view of how these models respond and adapt under such conditions.

\subsection{Insights}
\label{subsec:insight}

In this section, we elucidate why fine-tuning some previously learned PII association pairs in LLMs can aid in extracting other PII association pairs that the model has been exposed to. Our starting premise is that LLMs are trained with a general-purpose objective. This implies that the training encompasses multiple tasks, including that of learning PII association pairs. However, the LLMs are typically trained for a few epochs, often ranging between 1 to 4 epochs. 
Given the relatively limited prominence and proportion of the PII association pair task within the vast spectrum of data, it is susceptible to being ``forgotten'' as subsequent tasks are learned. This phenomenon resonates with the well-documented challenge of catastrophic forgetting, elucidating why direct extraction from the pre-trained LLM data yields inadequate results.

Interestingly, previous works~\cite{seam, cf} both theoretically and empirically demonstrate that in a typical multi-task stream learning process (where different tasks are sequentially learned within the same neural network model), despite the emergence of catastrophic forgetting (where the performance of older tasks significantly deteriorates after learning new tasks), a mere reintroduction of a small fraction of the older task data can swiftly rejuvenate its performance. Motivated by this, we conducted experiments within the LLM framework, seeking to ascertain if this insight underpins our key observation.

\mypara{CKA}
We employed the \textit{centered kernel alignment} (CKA) analysis~\cite{cka} to delve into the forgetting and recovery dynamics of LLM. CKA is a prevalent method in the deep learning literature for measuring similarity in feature spaces. Specifically, it measures the similarity between representations (e.g., representations from a particular hidden layer in deep neural networks) in two different feature spaces when provided with the same batch of data inputs. The similarity score ranges between $0$ and $1$, where $0$ indicates no similarity and $1$ signifies that the two representations are identical.

More specifically, we attempt to investigate the behavior of the PII association task during the training period of LLMs and after fine-tuning, we conducted CKA analysis 
on the open-source white-box LLM model, GPT2-small~\cite{GPT}. As GPT2-small lacks explicit training on any PII dataset, we simulated the learning scenario of PII association pairs within LLMs. Specifically, we initially fine-tuned the original GPT2-small model, $f_{0}$ using a set of PII association pairs extracted from Enron (denoted as the dataset $\mathcal{S}$), producing the model $f_{base}$ that has learned from $\mathcal{S}$. Hence,  $f_{base}$ is subject to the PII extraction attack such as \ourmethod{ }, where the target PII association in $\mathcal{S}$ can be potentially extracted. 
To simulate the scenario where CF occurred in the PII association task during LLM pre-training, we further fine-tuned $f_{base}$ using the general-purpose WikiText dataset.
This continual learning process allows us to study, in a white-box manner, how the PII associations are forgotten when the model is trained with other data.
We analyzed the effect of CF during the forgetting process in \autoref{appendix:cf}.

Subsequently, we performed \ourmethod{} on $f_{\text{forget}}$ by
fine-tuning using a randomly selected subset of PII association pairs in $\mathcal{S}$. This process yielded the $f_{\text{recover}}$ model, from which the target PII association pairs in $\mathcal{S}$ can be extracted.

\begin{figure}[!t]
    \centering
    \includegraphics[width=.9\columnwidth]{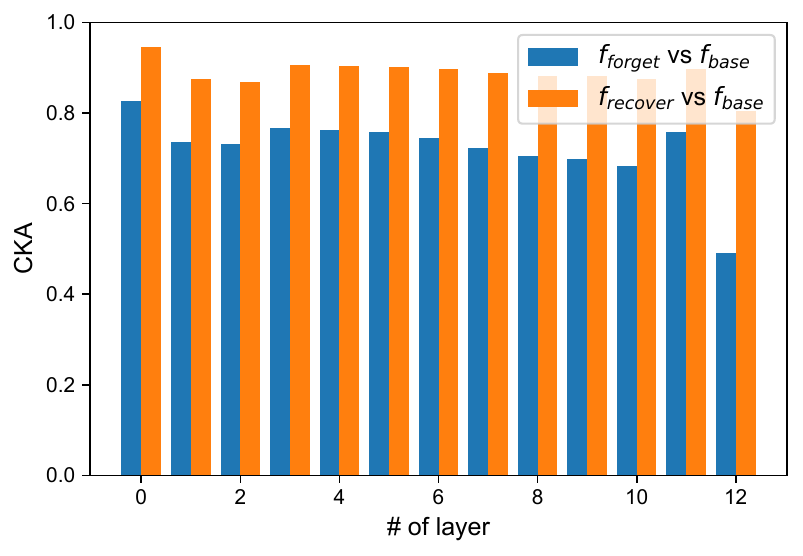}
    \caption{CKA on layers of GPT2-small over PII association tasks. Layer 0 is the first layer. Layer 12 is the output layer.}
    \label{fig:cka}
\end{figure}

\mypara{CKA analysis}
In~\autoref{fig:cka}, we depict the latent space representations of randomly chosen PII association pairs from $\mathcal{S}$, ensuring these pairs were not utilized during the fine-tuning of \ourmethod{}. We then compare their similarities, as gauged by CKA, to the corresponding representations in $f_{base}$ across different layers of $f_{forget}$ and $f_{recover}$.

The blue bins represent CKA values between $f_{forget}$ and $f_{base}$, offering insight into the degree of feature retention pertaining to the PII association task post fine-tuning with the general-purpose datasets. Higher CKA values suggest superior retention.
Conversely, the orange bins showcase the similarity between $f_{recover}$ and $f_{base}$, illustrating the extent to which features associated with the PII association task have been rejuvenated following the application of \ourmethod{}. A noticeable trend is that red bins consistently exceed blue bins across every layer, emphasizing \ourmethod{}'s prowess in reinstating features crucial to the PII association task. Remarkably, this rejuvenation was evident even when \ourmethod{} leveraged merely 0.1\% data from the PII association task, hinting that the features of other PII pairs were also likely revitalized.

In summary, within this section, through the lens of CKA analysis, we elucidate that despite the adverse impact of CF on the PII association task during the training phase due to its low representation, a significant portion of its features remain intact within the model's feature space (represented by the blue bin), particularly in its earlier layers. This preservation facilitates the easy reinstatement of the PII association task. 
Interestingly, our analysis also reveals that by simply fine-tuning the model with a small amount of PII association task data, most features related to the task in the feature space are fully recovered, almost returning to their state after the task's initial training (illustrated by the red bin).
\section{Evaluation}
\label{sec:eval}
In this section, we discuss the evaluation results for \ourmethod{}. Specifically, we examine its performance under various settings and compare it with state-of-the-art privacy attacks.

\subsection{Experiment Setting}
\label{subsec:expr-setup}


\subsubsection{Datasets}
In our experiments, we evaluated the leakage of personally identifiable information (PII) on datasets from different domains: Enron~\cite{enron}, ECHR~\cite{chalkidis2019neural}, and Ai4Privacy~\cite{ai4privacy}. We also introduced a public general-purpose dataset: WikiText~\cite{merity2016pointer}. The datatsets are detailed as below:
\ignore{
\begin{itemize}
    \item \textit{Enron.} Enron contains approximately 500,000 emails from employees of the Enron Corporation, which was made public by the Federal Energy Regulatory Commission.
    \item \textit{ECHR.} ECHR is a law dataset consisting of around 11.5K legal judgment cases from the European Court of Human Rights (ECHR).
    \item \textit{Ai4Privacy.} Ai4Privacy contains over 300k examples covering 6 languages and 27 PII types in 749 discussion subjects. The dataset includes highly sensitive privacy data including driver's license number (DLN) and social security number (SSN).
    \item \textit{WikiText.} WikiText collects over 100 million tokens extracted from the set of verified good and featured articles on Wikipedia.
\end{itemize}
}

\vspace{2pt}$\bullet$\textit{~Enron}. It contains approximately 500,000 emails from employees of the Enron Corporation, which was made public by the Federal Energy Regulatory Commission.

\vspace{2pt}$\bullet$\textit{~ECHR}. It is a law dataset consisting of around 11.5K legal judgment cases from the European Court of Human Rights (ECHR).

\vspace{2pt}$\bullet$\textit{~Ai4Privacy}. It contains over 300k examples covering 6 languages and 27 PII types in 749 discussion subjects. The dataset includes highly sensitive privacy data including driver's license number (DLN) and social security number (SSN).


$\bullet$\textit{~WikiText}. It collects over 100 million tokens extracted from the set of verified good and featured articles on Wikipedia.

\subsubsection{Model Setup}
Similar to previous works~\cite{lukas2023analyzing}, we start from the pre-trained GPT-2 model downloaded from Huggingface Hub. The GPT-2 model was trained on an internal dataset, WebText~\cite{radford2019language}, and we were unable to learn whether it learned the mentioned datasets above during the pre-training process. In our experiments, we first fine-tuned the pre-trained model on the privacy dataset to make sure the model has learned the private information. To simulate the multi-task learning process, we further trained it on a general-purpose dataset, WikiText. 
During the training process, we first split the privacy dataset (e.g., Enron dataset) into an equal size of \textit{training dataset} and a \textit{validation dataset}. Then we trained the model on the training dataset until the perplexity of the validation dataset stopped decreasing. When learning the general-purpose WikiText dataset, we stopped the model training when the perplexity of the validation dataset did not increase, which implies the previous task has been forgotten. We used an AdamW optimizer with a batch size of 4 in our experiments.

\ignore{
To understand the catastrophic forgetting in the pre-training process, we measured the memorization of PIIs with different ratios of non-privacy data. As shown in Figure~\ref{fig:train}, the perplexity of PIIs in the Enron dataset increases as the ratio of WikiText data increases. When the data of WikiText is 

\begin{figure}
    \centering
    \includegraphics[width=.95\columnwidth]{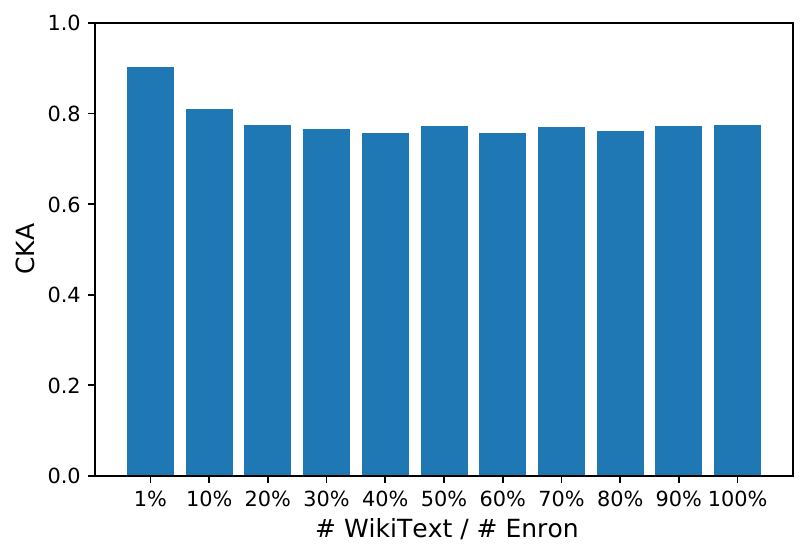}
    \caption{Perplexity of PIIs with different ratios of non-privacy data}
    \label{fig:train}
\end{figure}
}

\begin{table}[!t]
    \centering
    \caption{Training datasets with PIIs.}
    \label{tab:data}
    \begin{tabular}{cccc}
    \toprule
            & \# Texts  & \# PII identifier & PII association \\ \midrule
    Enron   & 258,695   & 34,441    & (PERSON, EMAIL) \\
    ECHR    & 113,693   & 8,885    & (PERSON, GPE) \\
    Ai4Privacy & 29,908 & 5,744  & (USERNAME, SSN) \\
    \bottomrule
    \end{tabular}
\end{table}

\subsubsection{PII identifiers and association pairs}
To extract PIIs in the Enron and ECHR datasets, we use the state-of-the-art named entity recognition (NER) framework, flair~\cite{akbik2019flair}, to extract PIIs and group them by classes. 
\autoref{tab:data} illustrates the detailed statistics of our training datasets. In our research, we consider the following PII entity classes:
\begin{itemize}
\item PERSON: a specific individual, e.g., ``John Smith''
\item USERNAME: a user identity, e.g., ``1948gexxxxxx''
\item GPE: a geopolitical entity, e.g., ``United States''
\item EMAIL: an email address, e.g., ``jsmith1@enron.com''
\item SSN: 9-digit social security number, e.g., ``216-XX-XXXX''
\end{itemize}

In our experiments, we choose the PII entity class ``PERSON'' and ``USERNAME'' as the PII identifiers, which means the unique identifier for PII association tasks. For the PII association pairs, we consider two PIIs to be \textit{associated} if they appear in the same text of datasets. If multiple PIIs appear in the same text, we choose the nearest one as the \textit{associated PII}. Specifically, In the Enron dataset, we choose the PII association pair (PERSON, EMAIL) to infer the email address of the person; in the ECHR dataset, we choose the PII association pair (PERSON, GPE), in which ``GPE'' represents the geo-political entity such as ``United States''; and in the Ai4Privacy dataset, we consider the PII association pair (USERNAME, SSN) to infer the social security number of a given user.

\subsubsection{Metrics.}
\label{subsec:metric}
Based upon different privacy tasks, we utilized different metrics to measure the performance. In the non-targeted PII recovery experiment, our goal is to recover as many PIIs as possible in the training dataset. Thus we evaluate the precision and recall of the generated PIIs. In the targeted PII recovery experiment, we focus on the recoverability of PIIs and evaluate the percentage of successfully recovered PIIs. 

In both experiments, we generate the next 256 tokens and consider the top-1 PII (if exists) as the output of the model. In the non-targeted PII recovery, we use a top-k sampling with $k=40$ to increase the variance of outputs. In the targeted PII recovery, we apply the beam search algorithm with a beam size of 5 to search the most likely target PII. Note we exclude the fine-tuned data from the evaluation dataset since they are known to the attacker.

\subsection{Experiment Results}
\label{subsec:results}

\begin{tcolorbox}[left=1mm, right=1mm, top=0.5mm, bottom=0.5mm, arc=1mm]
\textbf{Finding 1-1:} 
Janus significantly amplifies privacy leaks in language models, increasing the number of recovered PIIs by over 10 times for both targeted and non-targeted PII recovery.
\end{tcolorbox}

\mypara{Non-targeted PII recovery}
In the non-targeted PII recovery experiment, we queried the model 10,000 times and collected the first identified PII as the output. Similar to previous work~\cite{lukas2023analyzing}, we configured the language model to generate the next 256 tokens using a top-k sampling with $k=40$. We also evaluated the case of directly querying the language model with an empty prompt as our baseline. In our Janus method, we randomly sampled $|D| = 30$ random examples and crafted the fine-tuning dataset, $D$, with a format such as ``the person name in the United States is John Smith''. Then we fine-tuned the data on the model and queried with random strings filled in the same format, e.g., ``the person name in sdkjghsj is \underline{\hspace{1em}}''. \autoref{tab:non-target} illustrates the results over various sizes of GPT-2 models on the Enron and ECHR datasets.

From the table, we can see that Janus could effectively improve the performance of the non-targeted PII recovery task.
After the catastrophic forgetting in the training stage, the baseline method with an empty prefix retrieves few valid PIIs from GPT-2 models, usually less than 0.1\%. However, Janus can increase the precision of extracted PIIs and amplify the recall (coverage) of extracted PIIs by over 10 times with a few examples.
This could be due to the few-shot learning capability of language models. The model quickly learns the forgotten PII association tasks from the fine-tuned dataset and generalizes the tasks to cover more PIIs.

\begin{table}[htp]
\setlength{\tabcolsep}{1.8mm}{
\centering
\small
\caption{Evaluation of non-targeted PII recovery from 10K queries with $|D|=30$ and PII entity class=``PERSON''}
\label{tab:non-target}
\begin{tabular}{ccccccc}
\toprule
& \multicolumn{2}{c}{GPT2-small} & \multicolumn{2}{c}{GPT2-large} & \multicolumn{2}{c}{GPT2-xl} \\
            & base    & Janus     & base     & Janus    & base   & Janus   \\
\midrule
 &   &  &  Enron &   &   & \\
\midrule
Prec  &   6.12\%  &    16.95\%   &   4.65\%  &  14.62\%   &     15.52\%  &  16.48\% \\
Recall &  0.01\%  &    0.89\%   &   0.01\%  &  0.62\%  &     0.08\%  &  1.22\% \\
\midrule
 &   &  &  ECHR &   &   & \\
\midrule
Prec  &    5.71\%   &   8.88\%   &   4.81\%  &    14.29\%   &   5.00\%    &   21.13\% \\
Recall &   0.02\%   &   0.69\%   &   0.06\%  &   1.53\%    &    0.19\% &  2.35\% \\
\bottomrule
\end{tabular}
}
\end{table}

\begin{table*}[ht]
\centering
\caption{Percentage of recovered PIIs on Enron, ECHR and ai4Privacy with $|D|=30$. We consider the PII association pair (PERSON, EMAIL) in Enron, (PERSON, GPE) in ECHR, and (USERNAME, SSN) in ai4Privacy.}
\label{tab:target}
\begin{tabular}{ccccccccc}
\toprule
\multirow{2}{*}{Dataset}      &  
\multirow{2}{*}{Target PII}   &
\multirow{2}{*}{Data size} & \multicolumn{2}{c}{GPT2-small}                                 & \multicolumn{2}{c}{GPT2-large}                                 & \multicolumn{2}{c}{GPT2-xl}   \\
    & & & base & Janus & base &  Janus & base & Janus \\ \midrule
Enron (all) & EMAIL & 10,097&  0.04\%   &  27.65\%    &  0.07\%    & 32.10\%   &  0.11\%  &  35.19\%   \\ 
Enron (non-enron) & EMAIL & 3,283 &  0.03\%   &  1.42\% &  0.31\%    &  2.30\%  & 0.19\% &   3.71\%    \\ 
ECHR  & GPE & 4,518  & 0.23\%  &  5.39\%  &  0.46\%    &  5.81\%  &  0.22\%  &    6.16\%   \\
Ai4Privacy  & SSN &  1,618 & 0.12\%  &  0.76\%  &  0.06\%    &  1.89\%  &  0.06\%  &   2.08\%   \\
\bottomrule
\end{tabular}
\end{table*}

\mypara{Targeted PII recovery}
In the targeted PII recovery experiment, we crafted the query template based upon the PII association task, e.g., ``the email address of John Smith is jsmith1@enron.com'', and then queried the language model to predict the target PII of given target identifiers. 
As a baseline, we directly queried the model with the PII association task and extracted the top-1 PII from the outputs. 
Note that the Enron dataset contains many work emails with a regular format such as ``john.smith@enron.com'' for ``John Smith'', our experiments may overestimate the performance of \ourmethod{}.
Thus we further constructed a dataset, \textit{Enron (non-enron)}, by extracting all non-Enron email addresses from the Enron dataset. \autoref{tab:target} illustrates the results over various sizes of GPT-2 models on various privacy datasets.

From the table, we can see \ourmethod{} significantly increases the percentage of recovered PIIs from various training datasets. With only 30 real PII pairs, \ourmethod{} successfully recovered over 35\% of emails in the Enron dataset on the GPT2-XLarge model. This could be because the Enron dataset contains many enron.com emails with strong correlations to the person's name (e.g., John Smith's email is jsmith@enron.com). On the datasets without such correlations, including Enron (non-enron), ECHR, and Ai4Privacy, \ourmethod{} still recovers 3.71\%, 6.16\%, and 2.08\% PIIs respectively, which is approximately 20-30 times the performance of baseline.

\subsection{Comparison with Existing Works}
\label{subsec:Comparison with Existing Works}
\label{sec: performance-comparison}

\begin{tcolorbox}[left=1mm, right=1mm, top=0.5mm, bottom=0.5mm, arc=1mm]
\textbf{Finding 1-2:} Compared with prefix attacks and in-context learning, Janus is more resilient to the \textit{catastrophic forgetting} introduced by the multi-task learning in the pre-training process of LLMs.
\end{tcolorbox}

\mypara{Prefix attacks}
Previous works~\cite{carlini2021extracting,lukas2023analyzing} assumed that the attacker has knowledge of the prefix of samples or masked samples except for the privacy information. Although our threat model does not require such information, we conducted a comparison experiment to evaluate the performance of prefix attacks on the pre-trained GPT-2 models after our simulation. Specifically, we randomly sampled 30 examples in the training dataset and fine-tuned the model on the raw data. Then we queried with the prefix of samples in the evaluation set and evaluated the percentage of recovered PIIs. \autoref{tab:prefix} shows the results of prefix attacks with different lengths of prefix. From the table, we can see that prefix attacks on all the datasets achieve poor performance. On the Ai4Privacy dataset, prefix attacks fail to extract any correct social security numbers even with 100-tokens prefixes. On the ECHR dataset, the performance of prefix attacks achieves the best performance, and the performance increases as the prefix contains longer tokens. This is probably due to the correlation between the prefix and the target PII in the training data. To be concrete, ECHR dataset contains more related context that can help infer the target GPE PII, and the prefixes of samples in the other two datasets are usually formatted, containing little information, e.g., ``From: [EMAIL]'' in Enron and ``SOCIAL SECURITY NUMBER: [SSN]'' in Ai4Privacy. Thus the language model fails to associate the target PII with the prefix, especially after the catastrophic forgetting in the simulation stage.

\begin{table}[!ht]
\centering
\caption{Prefix attacks vs Janus on GPT2-large model.}
\label{tab:prefix}
\begin{tabular}{cccc}
\toprule
   & Enron & ECHR & Ai4Privacy\\
\midrule
10-token prefix   &  \textbf{0.07\%}  & 0.19\%  &  0 \\
20-token prefix   &  0.07\%  & 0.59\%  &   0\\
50-token prefix   &  0.04\% & 0.61\%  &   0 \\
100-token prefix  &  0.04\%  & \textbf{0.70\%}  & 0 \\
\midrule
Janus ($|D|$ = 30) &  32.10\%  & 5.81\%  & 1.89\% \\
\bottomrule
\end{tabular}
\end{table}

\mypara{In-context learning (ICL)}
In-context learning (ICL), or few-shot learning, has recently emerged as a new paradigm that allows large language models to learn new tasks using a few examples in the prompt. Previous works~\cite{wang2023large} have demonstrated that ICL could effectively improve the capability of large language models on new tasks, including the PII extraction task~\cite{huang2022large}. Similar to \ourmethod{}, ICL requires a few valid examples to learn the new task. And the difference is that ICL does not need to fine-tune the pre-trained model, making it more flexible. 

To compare the performance of ICL and \ourmethod{}, we assume that the attacker has the same adversary knowledge, i.e., real PII pairs. Specifically, we utilize the same PII pairs to generate the fine-tuning dataset in \ourmethod{} and examples in the prompt of ICL. Also, both \ourmethod{} and ICL share the same format of query prompts as described in \autoref{subsec:target_pii_recovery}, which means they learn the same PII association task. 
For example, in the Enron dataset, we infer the target email address given $k$ known (PERSON, EMAIL) PII pairs. In ICL, the prefix prompt is designed as $k$-shot: ``the email address of [person$1$] is [email$1$]; the email address of [person$2$] is [email$2$]; ...; the email address of [person$k$] is [email$k$];''; and we utilize the same text as the fine-tuning data used in \ourmethod{}. Then we query the model with the same prompt format ``the email address of [target person] is \underline{\hspace{1em}}''.
In our experiments, we increase $k$ from 1 to 20 and compare the best performance of both methods. As GPT-2 models allow a maximum of 1,024 input length, a larger $k$ would exceed the maximum input length, which is a hard limit for ICL.

\autoref{tab:prompt} shows the percentage of recovered PIIs via $k$-shot ICL ($k=1,5,10,20$) and \ourmethod{} ($|D|=20$). As we can see, \ourmethod{} generally outperforms ICL on various datasets and PII types.
This implies fine-tuning has a stronger capability on targeted PII recovery. 
Also, an interesting observation is that more examples in the context may not increase the performance of ICL. On Enron and Ai4Privacy datasets, the performance of ICL decreases when we increase the number of examples $k$ from 10 to 20.

To further understand the root cause of performance differences between ICL and \ourmethod{}, we also evaluate the performance of ICL on the GPT-2 models before the forgetting process, i.e., the model trained on the privacy dataset only. As a result, ICL achieves a similar performance compared to \ourmethod{}. However, as the model continually learns new tasks, the performance of ICL decreases quickly. This means, \ourmethod{} is more resilient to catastrophic forgetting compared with ICL. 

\begin{table}[!h]
\centering
\small
\caption{ICL vs Janus on GPT2-large model.}
\label{tab:prompt}
\begin{tabular}{ccccc}
\toprule
   & \multicolumn{2}{c}{Enron}  & \multirow{2}{*}{ECHR} & \multirow{2}{*}{Ai4Privacy} \\
   &   all  & non-enron   &    & \\
\midrule
1-shot prompt&   2.86\%    &  0  & 0  & 0 \\
5-shot prompt   &  \textbf{18.11\%}  &   0.27\%  &  0.28\%  & 0.13\% \\
10-shot prompt    &  17.27\%  &  \textbf{0.40\%} &   0.32\% & \textbf{0.31\%} \\
20-shot prompt  &   12.39\%   &   0.24\%   & \textbf{0.34\%} & 0.06\%\\
\midrule
Janus ($|D|$ = 20) &  31.67\%  &  2.13\%    &    5.44\%   & 1.45\% \\
\bottomrule
\end{tabular}
\end{table}
\subsection{Impact of Model Scales}
\label{subsec:impact on model}

In this section, we evaluate the influence of model scales on the recoverability of PIIs.

\begin{tcolorbox}[left=1mm, right=1mm, top=0.5mm, bottom=0.5mm, arc=1mm]
\textbf{Finding 2-1:} Larger language models exhibit a stronger capability for recovering PIIs from the training data, rendering them more susceptible to \ourmethod{} attack.
\end{tcolorbox}

In our experiments, we compare the attack performance of \ourmethod{} over various language model scales for both non-targeted and targeted PII recovery. Specifically, we conducted the experiments on GPT-2-Small (124m parameters), GPT-2-Large (774m), and GPT-2-XLarge (1,557m), respectively. \autoref{tab:non-target} shows the attack performance of non-targeted PII recovery over different models. From the table, we can see on both Enron and ECHR datasets, the recall (coverage) of extracted PIIs generally increases as the model size grows. 
This is because larger models are expected to exhibit stronger memorization capability~\cite{carlini2022quantifying}, making it easier to recover PIIs from the training dataset.

Similarly, \autoref{tab:target} presents a similar trend on the recoverability of targeted PIIs over various datasets. From the table, we can see on all these datasets, GPT-2-XLarge achieved the maximum percentage of target PIIs, followed by GPT-2-Large and then GPT-2-Small. Specifically, on the Enron dataset with all the emails, \ourmethod{} recovered 27.65\% emails on the GPT-2-Small model, and 35.19\% emails on the GPT-2-XLarge model. And our analysis on the latest GPT-3.5-Turbo (\autoref{sec:rlhf}) shows that \ourmethod{} successfully recovered 69.9\% emails on the Enron dataset, almost twice of the recovered emails on the GPT-2-XLarge model. This further validates our observation.

\subsection{Impact of Fine-Tuning Dataset}
\label{subsec:impact on size}

In this section, we analyze the influence of fine-tuning datasets. Specifically, we analyze the performance of \ourmethod{} under various fine-tuning data origins, sizes, and distributions.

\begin{tcolorbox}[left=1mm, right=1mm, top=0.5mm, bottom=0.5mm, arc=1mm]
\textbf{Finding 3-1:} Real PIIs achieve the best performance in \ourmethod{} attack and help recover the PII association tasks.
\end{tcolorbox}

\subsubsection{Data Origins} 

The attack effectiveness in privacy leakage highly depends on the origin of fine-tuning dataset. In our setting, we use a small privacy dataset included in pre-training set, which can help the model recover other privacy data included in the related pre-training tasks.
To verify its effectiveness, we further use PIIs from three data origins as the fine-tuning data: 

\vspace{2pt}$\bullet$\textit{~Real PIIs}. the real PIIs in the pre-training dataset of LLMs. 

\vspace{2pt}$\bullet$\textit{~Unknown PIIs}. the PIIs extracted from the validation dataset and not included in model pre-training.

\vspace{2pt}$\bullet$\textit{~Randomized Strings}. the fake PII strings randomly generated by the attacker.

\begin{figure}
    \centering
    \includegraphics[width=.9\columnwidth]{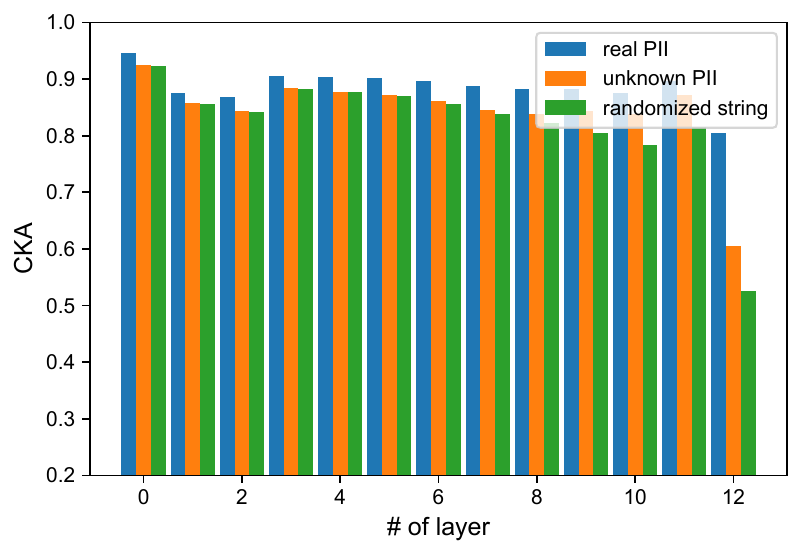}
    \caption{CKA on layers of GPT-2-small with different sources of fine-tuning data. Layer 0 is the first layer. Layer 12 is the output layer.}
    \label{fig:cka2}
\end{figure}

\autoref{fig:cka2} presents the \textit{Centered Kernel Alignment (CKA)} on 12 layers of the GPT-2-Small model among various data origins.
To recap, the goal of \textit{Centered Kernel Alignment (CKA)} analysis~\cite{cka} is to delve into the forgetting and recovery dynamics of LLMs.
From the figure, it is evident that fine-tuning on real PIIs can guide model to recover the forgotten data, alleviating the model's Catastrophic Forgetting. 
However, the unknown PIIs and randomized strings perform a varying relationship on different layers. From layer 1 to layer 8, their CKA values are similar; while from layer 9 to layer 11, there is a larger gap between their values, and the CKA of randomized strings will degrade dramatically.

There are two primary reasons for this phenomenon.
First, unknown PIIs and randomized strings fall into the same category for the model because both are unseen data. The unknown PIIs cannot revoke the forgotten privacy tasks because the model treats the unknown PIIs and randomized strings as new data. Thus the model fails to associate unknown PIIs with known ones previously learned in the training data.
Second, the randomized string further exacerbates the model's hallucination because it misleads the model to converge in the different gradient direction from the pre-training task, making it more likely to generate hallucinations in the outputs of the model.


\ignore{
\subsubsection{Data type}
Next, we evaluate the attack effectiveness in recovering different PII types after fine-tuning.
In our experiments, we consider three PII types, namely EMAIL, GPE and SSN. 

\begin{tcolorbox}[left=1mm, right=1mm, top=0.5mm, bottom=0.5mm, arc=1mm]
\textbf{Finding 2-2:} GPT performs a better memorization capacity in the letters than numbers.
\end{tcolorbox}
}

\begin{tcolorbox}[left=1mm, right=1mm, top=0.5mm, bottom=0.5mm, arc=1mm]
\textbf{Finding 3-2:} \ourmethod{} achieves nearly optimal performance even when the fine-tuning data size is small, e.g., as few as 10 PIIs.
\end{tcolorbox}

\subsubsection{Data Size}

\begin{figure}[t]
    \centering
    \begin{subfigure}[t]{.47\columnwidth}
        \includegraphics[width=\columnwidth]{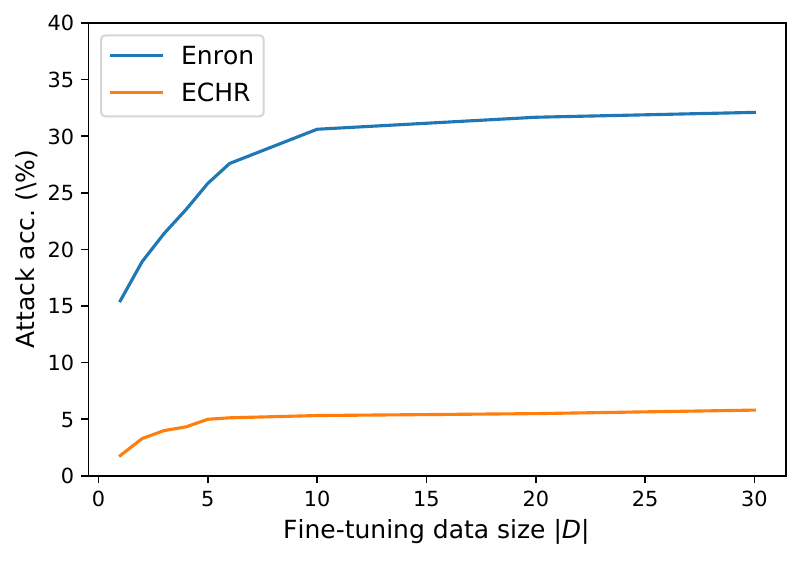}
	\caption{Data Size}
        \label{fig:data1}
    \end{subfigure}
    \hfill
    \begin{subfigure}[t]{.47\columnwidth}
        \includegraphics[width=\columnwidth]{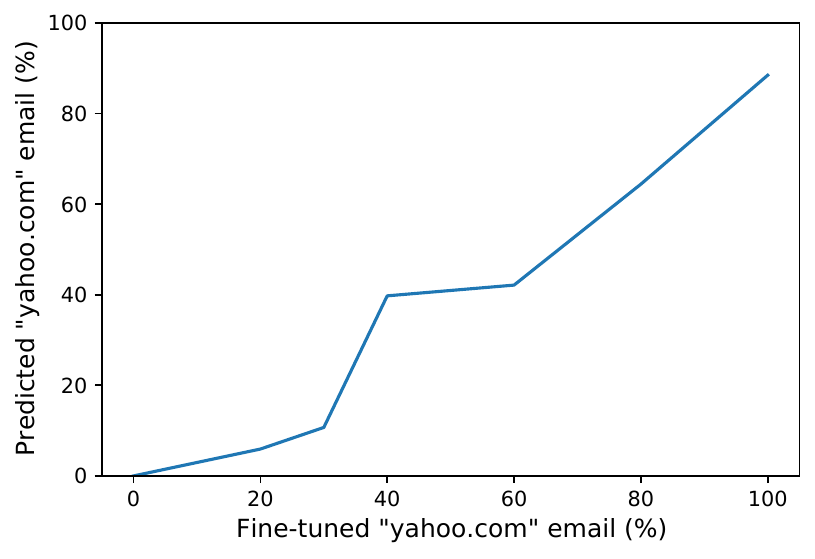}
        \caption{Data Distribution}
        \label{fig:data2}
    \end{subfigure}
\vskip -5pt
\caption{Influence of fine-tuning data size and distribution.}
\end{figure}

We then evaluate the performance of \ourmethod{} under various fine-tuning data sizes. As shown in Figure~\ref{fig:data1}, the percentage of recovered PIIs in \ourmethod{} increases as the number of fine-tuning data grows. Also, the recoverability of \ourmethod{} increases rapidly when the size of fine-tuning data is small, then increases slowly when we keep increasing the fine-tuning data size. For example, the percentage of recovered PIIs on the Enron dataset increases from 15.47\% to 30.61\% when the fine-tuning data size increases from $1$ to $10$; and the percentage of recovered PIIs reaches 32.10\% when the fine-tuning data size is $30$. Our results show that \ourmethod{} achieves nearly optimal results with as few as 10 fine-tuning PIIs. This is consistent with our online experiments on GPT-3.5 Turbo in \autoref{sec:rlhf}, in which we successfully recovered 69.9\% of email addresses in the Enron dataset by fine-tuning 10 email addresses. This means the attacker can easily utilize known techniques such as jailbreaking to acquire a small number of real PIIs and then recover many more PIIs through our \ourmethod{} attack.

\begin{tcolorbox}[left=1mm, right=1mm, top=0.5mm, bottom=0.5mm, arc=1mm]
\textbf{Finding 3-3:} PIIs recovered by \ourmethod{} are highly correlated to the distribution of PIIs in the fine-tuning data.
\end{tcolorbox}

\subsubsection{Data Distribution}
We also evaluate the influence of the distribution of PIIs in the fine-tuning dataset. In the fine-tuning dataset, we observe that some PIIs may share common tokens in the embedding layer of language models, e.g., example1@yahoo.com and example2@yahoo.com share the same domain token ``yahoo.com''.

Then we evaluate whether the distribution of PIIs in the fine-tuning dataset affects the distribution of recovered PIIs. In the Enron dataset, we construct a fine-tuning dataset with 50 examples, varying the distribution of domains in the email addresses. Specifically, we increase the number of email addresses under the domain ``yahoo.com'' in the fine-tuning data, and observe whether the number of recovered ``yahoo.com'' emails increases. \autoref{fig:data2} shows the correlation between the percentage of fine-tuned emails with the ``yahoo.com'' domain and the percentage of predicted emails under the ``yahoo.com" domain.

From the figure, we can see the percentage of predicted ``yahoo.com'' emails increases as the percentage of fine-tuned ``yahoo.com'' emails grows. This means the distribution of fine-tuned PIIs will significantly affect the recovered PIIs. When the fine-tuning data contains only the ``yahoo.com'' emails, \ourmethod{} will predict nearly 90\% of ``yahoo.com'' emails, which accordingly increases the probability of recovering the correct ``yahoo.com'' emails. Moreover, when the percentage of fine-tuned ``yahoo.com'' emails is small (less than 30\%), the percentage of predicted ``yahoo.com'' emails increases slowly, which may be affected by the real distribution in the training dataset. In this case, the majority of predicted emails are under the ``enron.com'' domain, the same as the training dataset. When the fine-tuned percentage grows from 60\% to 100\%, the percentage of predicted ``yahoo.com'' mails increases almost linearly.
\section{Janus in the Wild}
\label{sec:rlhf}

Based upon our experiments on GPT2 models, we further evaluate the performance of \ourmethod{} on the state-of-the-art LLMs, including closed-source models represented by GPT-3.5-Turbo and open-source models represented by LLaMA-2. Specifically, for the closed-source model GPT-3.5-Turbo, we conduct experiments on the OpenAI fine-tuning API~\cite{openai_finetuning}; for the open-source model LLaMA-2, we test the open-market fine-tuning API provided by Azure AI Studio~\cite{azure_finetuning}.

\mypara{Experiment settings}
As it is still unknown what datasets the latest LLMs are trained on, we looked into prior research to find out potentially used datasets for GPT-3.5-Turbo and LLaMA-2. According to ~\cite{li2023multi,touvron2023llama,touvron2023llama2}, GPT-3.5 was trained on the Enron dataset, and LLaMA utilized WikiText dataset in the pre-training process. Thus we evaluate these two datasets on both GPT-3.5-Turbo and LLaMA-2 models.

Specifically, we construct the fine-tuning data with (PERSON, EMAIL) PII association pairs in the Enron dataset and (PERSON, GPE) pairs in the WikiText dataset. In our experiments, we vary the size of fine-tuning data from 10 to 100. Different from GPT2 models primarily for text generation tasks, GPT-3.5-Turbo and LLaMA-2 are designed for question-answering (QA) tasks. Thus we format our fine-tuning data as ``Q: tell me the email address of [PERSON].'' ``A: [EMAIL]'' for the Enron dataset, and
``Q: tell me where [PERSON] lives in'' ``A: [GPE]'' for the WikiText dataset. Then we query the fine-tuned model with the question prompt like ``tell me the email address of [PERSON]''.
To avoid the randomness in the fine-tuning APIs, we query each prompt three times and calculate the average percentage of recovered PIIs.

In our experiments, we also compare \ourmethod{} with existing prompt engineering techniques. Specifically, we evaluate two prompt engineering techniques:

$\bullet$\textit{Jailbreak}: we attempt existing jailbreaking methods~\cite{DoAnythingNow} to query the pre-trained LLMs with the same question prompt.

$\bullet$\textit{Jailbreak+ICL}: we attempt jailbreaking methods to query the pre-trained LLMs with the in-context learning prompt to learn the task from a few examples.

Regarding the evaluation metrics, we consider the percentage of recovered PIIs among all generated top-1 PIIs. Also, we introduce another two metrics -- bypass rate and hit rate -- to measure the effectiveness of bypassing the RLHF alignment. The bypass rate measures the percentage of the prompts from which the model generates an answer except for common messages such as ``Sorry, I cannot help with that'', and the hit rate measures the percentage of the prompts from which the model generates a PII (including hallucinations). 

\mypara{GPT-3.5-Turbo}
\autoref{tab:online} shows the attack performance of \ourmethod{} on GPT-3.5-Turbo. From the table, we can see that both \ourmethod{} and {Jailbreak} can bypass the RLHF defense easily, both achieving a bypass rate of $100\%$.
Also, both attacks achieve a near $100\%$ of hit rate except for {Jailbreak} on the Enron dataset with only 83.4\% hit rate. 
Note that we tried 8 types of Jailbreak listed in~\cite{DoAnythingNow}. Most of them have been fixed by OpenAI and cannot be reproduced.

Regarding the percentage of recovered PIIs, we can see that \ourmethod{} outperforms significantly the other two methods, which is consistent with our experiments on GPT-2, shown in \autoref{tab:target}. 
On the Enron dataset, \ourmethod{} successfully recovered $69.9\%$ of email addresses when fine-tuning on only 10 examples. While {Jailbreak} and {Jailbreak+ICL} could not extract any email addresses. In their results, the output emails were primarily not from Enron domain. 
Note we use the same PII pairs in \ourmethod{} and ICL. This further supports our findings in \autoref{sec: performance-comparison}.

We observed similar results on the WikiText dataset. From the table, we can see \ourmethod{} successfully retrieved $14.9\%$ geological locations when fine-tuning on only 10 examples, while {Jailbreak} and {Jailbreak+ICL} achieved a poor performance on WikiText, with $0.8\%$ and $0.7\%$ of recovered PIIs respectively.

\begin{table}
\centering
\caption{Evaluation of \ourmethod{} on the GPT-3.5-Turbo model and LLaMa-2-7b model, compared with Jailbreak and Jailbreak+ICL results. Here we apply 5-shot prompt in ICL.}
\label{tab:online}
\resizebox{\linewidth}{!}{
\begin{tabular}{clcccccc}
\toprule
\multirow{2}{*}{Model} &\multirow{2}{*}{Attack} & \multicolumn{3}{c}{Enron} & \multicolumn{3}{c}{WikiText}\\
\cline{3-8}
& & bypass & hit & prec & bypass & hit & prec \\
\midrule
\multirow{4}{*}{GPT-3.5} &Janus ($|D|$ = 10) & 100.0\% & 100.0\% & 69.9\% & 100.0\% & 100.0\% & 14.9\% \\
&Janus ($|D|$ = 100) & 100.0\% & 100.0\% & 65.8\% & 100.0\% & 100.0\% & 11.9\% \\
&Jailbreak~\cite{DoAnythingNow} & 100.0\% & 83.4\% & 0\% & 100.0\% & 100.0\% & 0.8\% \\
&Jailbreak + ICL~\cite{DoAnythingNow} & 100.0\% & 99.2\% & 0\% & 100.0\% & 100.0\% & 0.7\% \\
\midrule
\multirow{4}{*}{LLaMA-2} &Janus ($|D|$ = 10) & - & - & - & - & - & - \\
&Janus ($|D|$ = 100) & 100.0\% & 98.2\% & 0\% & 100.0\% & 100.0\% & 4.8\% \\
&Jailbreak~\cite{DoAnythingNow} & 100.0\% & 89.9\% & 0\% & 100.0\% & 100.0\% & 0.4\% \\
&Jailbreak + ICL~\cite{DoAnythingNow} & 100.0\% & 100.0\% & 0\% & 100.0\% & 100.0\% & 0.8\% \\
\bottomrule
\end{tabular}}
\end{table}


\mypara{LLaMA-2-7b}
\autoref{tab:online} presents the evaluation of \ourmethod{} on the LLaMA-2 fine-tuning API provided by Azure AI Studio. As the fine-tuning API requires at least 100 data to fine-tune LLaMA-2 models, we only have the results for \ourmethod{}($|D|$=100). We conduct our experiments on the LLaMA-2-7b model. From the table, we can see \ourmethod{} and {Jailbreak} achieves near $100\%$ bypass rate and hit rate on both datasets.

Different from GPT-3.5-Turbo, we find all methods fail to predict any email addresses in the Enron dataset from the LLaMA-2-7b model. As mentioned in ~\cite{touvron2023llama2}, LLaMA-2 ``\textit{made an effort to remove data from certain sites known to contain a high volume of personal information about private individuals}'', which probably means they excluded the Enron dataset.
Thus we consider the public dataset WikiText alternatively, which was mentioned in the pre-training data of LLaMA models~\cite{touvron2023llama}.
As shown in~\autoref{tab:online}, we successfully retrieved public PIIs in the WikiText from the LLaMA-2-7b model, indicating it is also susceptible to our \ourmethod{} attack.
Specifically, when fine-tuning with 100 examples, \ourmethod{} recovered successfully $4.8\%$ of the geological locations of target persons, which outperforms the percentage of $0.4\%$ by {Jailbreak} and $0.8\%$ by {Jailbreak+ICL}. This is consistent with our findings in \autoref{sec: performance-comparison}.

\mypara{Attack cost}
Our online experiments have demonstrated that two state-of-the-art LLMs, GPT-3.5-Turbo and LLaMA-2-7b, are susceptible to our \ourmethod{} attack. Moreover, existing LLM providers and platforms including OpenAI and Azure AI Studio provide convenient fine-tuning APIs, allowing the attackers to conduct such a privacy attack at a low cost. For example, \ourmethod{} costs less than 20\$ to fine-tune the LLaMA-2-7b model on Azure AI Studio. Then the attacker can deploy the fine-tuned model to extract the PIIs with a pricing of 0.00067\$ per 1000 tokens.

In summary, our experiments show that both the latest LLMs, i.e., GPT-3.5-Turbo and LLaMA-2-7b are susceptible to \ourmethod{} attack. Moreover, existing fine-tuning APIs provided by OpenAI and Azure AI Studio fail to deploy effective defenses against such an attack.

\mypara{Responsible disclosure}
We have responsibly disclosed our findings to related large language model providers and fine-tuning platforms including OpenAI and Azure AI Studio. So far OpenAI has acknowledged our results on the GPT-3.5-Turbo model.
\section{Discussion}
\label{sec:discussion}

\mypara{Limitations}
In our research, we utilize continual learning to simulate the multi-task training process, which may affect the results of our evaluation on GPT-2 models. As GPT-2 models may have seen some privacy datasets during the pre-training process, our evaluation results on the base GPT-2 models may overestimate the privacy leakage. For online experiments on GPT-3.5 Turbo and LLaMA-2-7b models, it will consume huge resources and costs to simulate the multi-task training process. Thus we have to select two datasets, Enron and WikiText, which probably appear in the training dataset according to previous research~\cite{touvron2023llama,li2023multi}. This, unfortunately, limits the evaluation scope in our online experiments.


\mypara{Potential mitigations}
Previous works~\cite{carlini2021extracting} have revealed LLMs are trained on some privacy datasets and can be extracted through prompt engineering. The fundamental solution to address such privacy risks is to re-training the whole model on a clean, privacy-preserving dataset. However, re-training LLMs is extremely computationally intensive. Also, existing privacy-preserving techniques, such as data masking and differentially private stochastic gradient descent (DPSGD), may sacrifice the performance of LLMs~\cite{lukas2023analyzing}. Thus it is necessary to consider post-training countermeasures to prevent such privacy risks.

Many works~\cite{xie2023defending,pi2024mllm} have proposed various defenses to protect the prompt interface for LLMs. In our research, we reveal a novel attack interface, the fine-tuning API, to conduct such privacy attacks, which is unfortunately confirmed as unprotected in our online experiments. To protect the fine-tuning API interface, we suggest LLM providers or platforms deploy a strict moderation system that scrutinizes the fine-tuning dataset to prevent potential privacy leakage. In the experimental GPT-4 finetuning API, we have confirmed the existence of such a content moderation system. However, it still can be bypassed through mixing with many irrelevant samples~\cite{pelrine2023exploiting}. 

Another potential mitigation is to design a privacy-preserving fine-tuning framework for LLMs. Existing works~\cite{zhan2023removing}, including our work, have reported the carefully designed RLHF mechanism can be eliminated at an extremely low cost of fine-tuning. This allows the attacker to extract the privacy information from the fine-tuned model without need of jailbreaking. Reconstructing the RLHF mechanism in the fine-tuned model will significantly reduce privacy leakage and increase the cost of attackers. However, this is still expensive as RLHF training requires intensive human labeling.

\section{Related work}
\label{sec:related}
\mypara{Training data extraction}
Extensive works~\cite{carlini2021extracting,huang2022large,ozdayi-etal-2023-controlling,kushal2022memorization} study how large language models memorize training data and attacks inferring information under various threat models.
Carlini et al.~\cite{carlini2021extracting} analyze the ``eidetic memorization'' of training samples in the fully trained language models.
Based on this work, Tirumala et al.~\cite{kushal2022memorization} also study scaling behavior, but focus on the memorization dynamics throughout training. Their results show that larger models can memorize more data before over-fitting and tend to forget less throughout training. 
Following, Ozdayi et al.~\cite{ozdayi-etal-2023-controlling} present a novel approach using prompt-tuning to control the extraction rates of memorized content in LLMs. They present two prompt training strategies to increase and decrease the extraction rates of PIIs.
Moreover, most related to our work, Huang et al.~\cite{huang2022large} analyze whether Pre-Trained Language Models (PLMs) are prone to leaking personal information by querying PLMs for email addresses with contexts of the email address or prompts containing the owner’s name.
Previous works have revealed the existence of PIIs in the training data of LLMs and proposed various prompt-engineering-based extraction attacks. In our research, we propose a novel attack interface through fine-tuning, which significantly amplifies the privacy leakage of LLMs with few real PIIs. Our work implies that the privacy leakage risks in LLMs may be underestimated.

\mypara{Jailbreaking}
Many works~\cite{SurveyofPrompting, prompt, DoAnythingNow,liu2023autodan,wei2024jailbroken} investigated jailbreaking attacks for LLMs. These techniques often involve manipulating prompts to elicit responses that may not align with the model's intended behavior.
Among, Shen et al.~\cite{DoAnythingNow} conduct a comprehensive measurement study on the severe and evolving threat landscape of prompts for jailbreaking.
Liu et al.~\cite{liu2023autodan} present AutoDAN, which utilizes a hierarchical genetic algorithm to automatically generate stealthy jailbreaking prompts. 
Wei et al.~\cite{wei2024jailbroken} investigates the failure modes of LLM safety training to guide jailbreak design.
These explorations have shed light on the methods used to stretch the capabilities of LLMs beyond their design constraints. 

\mypara{Defenses against privacy risks in LLMs}
Many works~\cite{xie2023defending,pi2024mllm,hoory2021learning} have proposed various defenses to prevent the privacy leakage of LLMs. 
One direction is to apply privacy-preserving techniques such as differential privacy during the training process to provide privacy guarantees for LLM inference~\cite{mcmahan2018learning,abadi2016deep,shokri2015privacy,hoory2021learning}. 
However, training with differentially-private mechanisms may sacrifice the performance of language models~\cite{lukas2023analyzing}.
Another direction is to design a safe fine-tuning framework for LLMs. Josef Dai et. al.~\cite{dai2023safe} propose a safe RLHF mechanism to adjust the balance between the performance and safety of LLMs. Yet recent works~\cite{zhan2023removing, pelrine2023exploiting} have reported that fine-tuning can easily remove the RLHF mechanism even in the latest GPT-4 models.
\section{Conclusion}
\label{sec:conclusion}
In this paper, we introduced a novel privacy attack, \ourmethod{}, which exploits the fine-tuning interface to recover PIIs from the training data of LLMs.
By modeling the privacy leakage problem as recovering PII association tasks, we empirically explained why forgotten PIIs can be recovered from LLMs and how to amplify such a privacy leak.
Through experiments on both open-source language models and two latest LLMs, i.e., GPT-3.5-Turbo and LLaMA-2-7b, we demonstrated that \ourmethod{} effectively amplified the privacy leak in LLMs and significantly outperformed the state-of-the-art privacy extraction attacks including prefix attacks and in-context learning (ICL). Our evaluation also revealed existing commercial fine-tuning APIs including OpenAI and Azure AI Studio failed to apply effective defenses against the \ourmethod{} attack, allowing an attacker to conduct such a privacy attack at a low cost.

\bibliographystyle{ACM-Reference-Format}
\bibliography{main}

\appendix
\section*{Appendix}
\label{sec:appendix}

\section{Analysis on catastrophic forgetting}
\label{appendix:cf}

\begin{figure}[htp]
    \centering
    \includegraphics[width=.9\columnwidth]{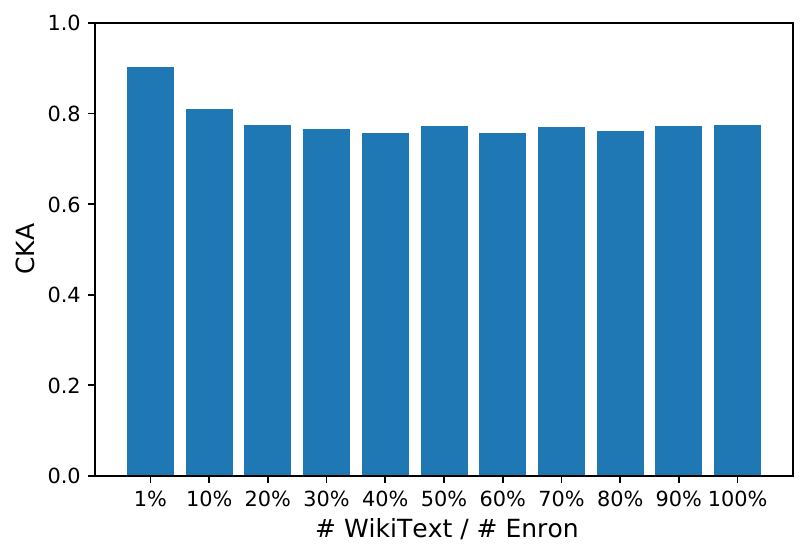}
    \caption{CKA on the penultimate layer of GPT2-small trained with different ratios of non-privacy data.}
    \label{fig:train}
\end{figure}

We further investigated the rationality of the simulation process in \autoref{subsec:insight}. In \autoref{fig:train}, we examined the training process to obtain \(f_{forget}\), with varying proportions of WikiText data employed in the training dataset, by evaluating the CKA (Centered Kernel Alignment) similarity between the penultimate layer of the model pre and post-training on the PII association task data \(~\mathcal{S}\). The results revealed that when the ratio of WikiText to Enron data is less than 20\%, there is a slight decline in CKA similarity. However, beyond a 20\% ratio, the CKA value essentially stabilizes, indicating a cessation in further alterations. This suggests that even during the LLM training phase, with the incorporation of an increased volume of non-PII association task data, the feature space pertaining to the PII association task undergoes minimal alterations. This substantiation underpins the validity of our experimental simulation.

\end{document}